# Diversity and Polarization of Research Performance: Evidence from Hungary

Sándor Soós[1], George Kampis[2]

[1]Institute for Research Policy Studies, Hungarian Academy of Sciences, Budapest, Hungary
[2]History and Philosophy of Science, Lorand Eötvös University, Budapest, Hungary
ssoos@colbud.hu

**Abstract** Measuring the intellectual diversity encoded in publication records as a proxy to the degree of interdisciplinarity has recently received considerable attention in the science mapping community. The present paper draws upon the use of the Stirling index as a diversity measure applied to a network model (customized science map) of research profiles, proposed by several authors. A modified version of the index is used and compared with the previous versions on a sample data set in order to rank top Hungarian research organizations (HROs) according to their research performance diversity. Results, unexpected in several respects, show that the modified index is a candidate for measuring the degree of *polarization* of a research profile. The study also points towards a possible typology of publication portfolios that instantiate different types of diversity.

## Introduction

In recent years, a number of measures have been proposed to formalize the notion of diversity in the context of research evaluation. A very promising direction of this investigation was the research experimenting with different diversity indices applied to publication records as proxies to interdisciplinarity. This approach successfully combined qualitative and quantitative dimensions of diversity, as it consisted of (1) developing a science map, (2) representing the publication record in terms of the map, and (3) subjecting the latter representation to quantitative analysis that was, in turn, sensitive to structural features of the map (cf. Rafols and Meyer 2010, 2007, Leydesdorff and Rafols 2010, Porter and Rafols 2009, Porter and Youtie 2009, Porter et al. 2007).

Inspired by this approach, here we propose a new version of the diversity index analized by the abovementioned authors, which was the Stirling index. As a use case, we apply this measure to the most prominent group of Hungarian research organizations, to compare the diversity of their publication performance along the modified and the original version of the Stirling index. Although the modification of the index is seemingly rather small, the obtained rank-order comparisons show that the proposed new use of the measure is capable of detecting hitherto unnoticed structural characteristics of research performance, as evaluated against the science maps picturing the publication profile of institutions.

We suggest that the presented use of the measure conveys an additional meaning to the concept of diversity. Based on its behavior, the modified index is discussed as a candidate for measuring the *polarization* of a research profile. Our results also point towards a possible typology of publication portfolios that instantiate different types of diversity.

**Related Work**

In the context of exploring the indicators of interdisciplinarity, in subsequent papers Porter, Rafols, Meyer and Leydesdorff have recently elaborated a formal apparatus for assessing the topical/intellectual diversity of a particular body of scholarly literature. The method is designed to be applicable to publication records. With some variation regarding the details of individual experiments, the methodology is constituted by the following common steps:

(1) A global science map is formed, against which the publication record can be evaluated. The map consists of a proximity network of subject categories (SCs) or journals included in the ISI–Thompson-Reuters databases (ISI or TR$^{TM}$), based on their respective citation patterns.
(2) A publication set P (similarly harvested from the TR$^{TM}$ databases) is projected onto the above map, using the fact that P can be expressed in terms of subject categories or journals to which the papers in P are assigned (in the TR$^{TM}$ databases). The result is a customized version of the global science map (or „basemap"), which offers a structural profile of P, thus depicting the number, weight and the relative position or distance of subject categories or journals present in P.
(3) On this structural profile of P diversity indices are imposed.

Among the set of diversity indices, the cited authors argued for various versions (i.e., different parametrizations) of the generalized Stirling index, a diversity measure of the following (simplified) form

$$\sum_{ij(i \neq j)} d_{ij} p_i p_j \, ,$$

where $d_{ij}$ is a distance value between elements $i$ and $j$ in the network (based on the degree of relatedness), multiplied by their respective contribution (relative share), $p_i$ and $p_j$. An attractive feature of this index is that is captures more aspects of the notion of diversity than most other indices around: it takes into account not just (1) the number and (2) the balance (distribution) of categories/journals constituting the publication record, but also the disciplinary relatedness of those elements. The rationale behind the notion is that the presence of categories distant in terms of the science map contribute more to the degree of diversity than the presence of closely related categories. That is, a research profile spread over a broader area of the map is indicated to be more diverse than a portfolio with the same number and the same distribution of categories, but mutually closely positioned in the network. In fact, the discussed feature is one of the main advantages of using a science map, i.e. a network of disciplinary descriptors, instead of relying on quantitative distributions of those descriptors. The general scheme described above has been realized in different ways by the authors who have been developing this apparatus. The approaches mostly differ in two respects: (1) the level of aggregation of the underlying science map, and—less strikingly—(2) the choice of the measure for the distance parameter in the Stirling index, $d_{ij}$. A typology of the existing approaches according to these dimensions is set out in Table 1.
To measure the interdisciplinarity (i.e., diversity) of a given research area, Rafols and Meyer utilized a map of individual papers, where the similarity network was built from the bibliographic coupling of those papers (Rafols and Meyer 2010). Rafols and Leydesdorff proposed to use the now-paradigmatic case of co-citation patterns of ISI Subject Categories (SCs) as the basemap depicting the current structure of science (Leydesdorff and Rafols 2009). SC maps has been used to measure a researcher's interdisciplinarity (through her publication list: Porter et. al 2007), or, as a demonstration, that of a publication record of an organization (Rafols–Porter–Leydesdorff 2009). The same group of authors, in measuring the interdisciplinarity of ISI journals, used a map constructed from the aggregated journal–journal citation matrix, i.e. a proximity map of the journals (Leydesdorff and Rafols 2010).

As to the variants of the Stirling index concerning more specifically the distance parameter used, most authors applied the cosine distance between two selected elements of the network. This choice was a direct consequence of the definition of the science map itself: a link between two elements was determined by their cosine similarity, i.e. the cosine measure for the two respective vectors. Since the weights of the edges were therefore proximity or closeness values, summing up those values straight away (as the index dictates) would result in a measure of „uniformity" (or Integration, as in the case of Porter et al 2007), instead of diversity. To express diversity, then, simply the complementer of cosine similarity was calculated for each link in the network, that is, the cosine distance: $d_{ij} = 1 - s_{ij}$, where $s_{ij} = \cos(i, j)$.

A conceptually different choice of the distance parameter can be seen in (Rafols and Meyer 2010). In measuring the diversity of a scientific discourse (that of bionanoscience, in the particular case), and the underlying map being the similarity network of papers in that topic, he experimented with the shortest path or the geodesic between two papers in the network. To avoid a pre-defined categorisation of papers, which is a pre-requisite for using the Stirling index, each paper was considered as a distinct category: as a result, the Stirling-index was reduced to the aggregated distances between the papers constituting the map. Formally, in this case, $d_{ij} = g_{ij}$, where $g_{ij}$ is the geodesic connecting $i$ and $j$, that is, the distance is the number of edges in the shortest path to get from paper $i$ to paper $j$.

**Table 1** *Typology of the Stirling index in measuring research diversity*

| | Formula (versions of the generalized Stirling index) | $d_{ij}$ | Underlying science map (level of aggregation) | Measuring diversity of… |
|---|---|---|---|---|
| *1* | $\sum_{ij(i \neq j)} d_{ij} p_i p_j$ | $1 - s_{ij}$, where sij=cos(i,j) | Similarity network of (1) journals (2) ISI Subject Categories (based on the cited and citing dimension) Rafols, Meyer, Porter, Leydesdorff | (1) journals, (2) work of researchers, (3) output of organizations |
| *2* | $\sum_{ij(i \neq j)} d_{ij}$ | $g_{ij}$ shortest path from i to j (# edges) | Similarity network of papers (based on bibliographic coupling) Rafols, Meyer | particular research area |

**Materials and Methods**

In what follows we endeavor to use and elaborate the Stirling index to measure the diversity of research. In applying the index our primary goal was to reveal and capture research diversity at the institutional level: as a use case, we address the case of Hungarian research organizations.

*The Hungarian Sample*

We compiled the publication record of Hungarian research organizations (HROs) for the recent decade, that is, covering the period 2000–2009. Data were retrieved from the TR[TM] databases through the ISI WoS portal. The resulting dataset was subjected to a thorough cleaning procedure, which consisted of the normalization of institutional names. Since institutional affiliations are represented in publications at various organizational levels (such as the university level or the faculty level), making organizations comparable required to aggregate publication entries at a selected, more-or-less uniform level. For a definite part of our data, organizations were referred to at the topmost level (e.g. MTA, Hungarian Academy of Sciences, an umbrella term for many research institutions) that could not be disaggregated, while others were cited at some lower levels (e.g. as some institution belonging to the MTA). Because of this feature of the dataset, and also to avoid imposing ad hoc hypotheses on the equivalence of organizational units, we used the top level for each organization. This resulted in a set of altogether 6154 research units, including Hungarian universities, governmental institutions concerned with research and development, and various companies exhibiting R&D activity.

In a subsequent step, this maximal list was reduced to a sample containing the „biggest" actors in Hungary, based on a ranking of the listed organizations according to the size of their publication record. In particular, actors were included that posessed a minimum of 100 publications per organization within the ten-year window of analysis. In the final set, 27 HRO were subjected to analysis. Organizations included in this sample are listed in Table 2 of the Appendix.

*Competence maps: mapping the research profile of the institutions*

The scheme of the mapping procedure, iterated for each targeted institution, consisted of three consecutive stages:

(1) Based on the journal–Subject Category assignments of the ISI databases, the publication record of the HRO was turned into a frequency distribution of Subject Categories, or, in other words, into a research profile.

(2) The research profile (or its normalized version) was projected on a basemap of science, representing the proximity of Subject Categories. This resulted in a network view of the HRO's profile, that we call a „competence map" of the particular institution.

(3) On the competence map, the modified Stirling index was calculated, providing a measure of diversity for the research portfolio.

The methodology described above is introduced by Rafols, Porter and Leydesdorff (2009) under the name of „overlay technique", since the essence of this approach is to create an overlay on the basemap of science, characteristic of the organization under study. The basemap, provided by the cited authors (see below) was, in this case, the similarity network of Subject Categories of ISI databases. Similarity is calculated from the SC–SC (asymmetrical) citation matrix. The relatedness of two SCs is detemined by their degree of co-citing the same SCs. This having been normalized, the similarity measure used was the cosine similarity of the citation vectors of individual SCs. The resulting science map is a weighted or valued network of Subject Categories, whereby the nodes are SCs, and the weighted edges between them represent similarity of the respecting SCs, the weights carrying the degree of similarity. In order to clarify the structure, edges under a certain threshold of similarity are omitted from the network (the empirical threshold was set to 0.15).

Projecting the research profiles of HROs onto this structure involved visualizing the share of individual SCs in the HRO's research profile on the map. In terms of visualization, the result is an altered map, where the size of a node is proportional to the share of the corresponding SC. The new map indicates the number, the proportion, and the relative position of Subject Categories in the research portfolio of the HRO in question. (Detailed methodology of the overlay technique can be found in (Rafols-Porter-Leydesdorff 2009).

The respective competence maps we generated for the 27 organizations of our final sample can be found in color at the URL http://www.hungarianscience.org/. The procedure was implemented in the R statistical software (R Development Core Team 2009), importing the basemap provided by Rafols and Leydesdorff (http://www.leydesdorff.net/overlaytoolkit/), using the Igraph package (Csardi and Nepusz 2006). The layout of the networks was calculated by the Fruchtermann-Reingold algorithm, placing more connected SCs in close proximity, while tossing less connected ones farther away. These visual representations of research portfolios are in good support of the quantitative measure of diversity, described in the next section.

*Altering the Stirling index*

Based upon the preparatory work exposed above, in the final step of our analysis, the modified Stirling index was calculated from the competence map of each institution. The basic idea behind the proposed form of the measure was to fully capture the properties of the underlying science map in evaluating the extent of diversity. In particular, the rationale for this experiment was to further explore the possibilities of using a network representation of the research profile instead of a quantitative distribution of publications over Subject Categories, and to utilize the structural properties of the map in as much as possible.

Our goal, in principle, was to formally capture the intuitive notion that the degree of diversity is closely linked to the „coverage" of the competence map, that is, essentially to the diameter of the underlying network. The diameter is determined by the shortest paths between any two Subject Categories: indeed, if, given a particular competence map, two SCs are close to each other in the sense that the number of steps (edges) leading from one SC to the other is relatively low, then, having these SC's in the research profile should not contribute to the degree of diversity. On the other hand, considerable work in distant areas, positioned for example at the opposite poles of the map in terms of path length, should maximize the measure for diversity.

Based on these considerations, we focused on the measure of path length as a candidate for the distance measure applied in the Stirling index: but, in contrast to Rafols's approach, who defined the distance parameter with the number of edges in the shortest path between the respective two SCs, we took into consideration the values (weights) of the edges in calculating the paths. That is, we used the sum of the cosine distances between SCs as the path length value. By doing so, we intended to incorporate not only the minimal number of steps connecting two SCs, but also the information on how „big" those steps are. In this way, we hoped to discriminate between cases where the same number of steps are required to reach areas in a close and in a much wider range of disciplinary similarities.

As a result, the version of the (generalized) Stirling index we used was the following:

$$\sum_{ij(i \neq j)} g_{ij}^{W} \, p_i \, p_j \text{ , where } g_{ij}^{W} = \text{sum of the weights of edges in the shortest path form } i \text{ to } j,$$

whereby the weights are the cosine distances between the respective SCs. An additional difference to the version used by Rafols is that here the relative share of categories (SCs) in the research profile also contributes to the end result.

It should be noted that utilizing path lengths as a distance measure is sensible partly due to the basemap not being a fully connected network. The frequency distribution of path lengths shows that in most cases it takes two steps (edges) to get from one SC to the other, at best (see Figure 4 in the Appendix). This feature of the basemap is partially yielded by the threshold imposed on the similarity measure, erasing a set of edges from the map. Given this setting, one might wonder how the newly introduced measure is related to the cosine distance, the measure most widely used in this context, and what difference it makes to use the valued path lengths instead of the original distance matrix of the network without thresholds on distances. These conceptual issues are briefly discussed in the Appendix of the paper.

In the next sections, we focus on the empirical results obtained when applying the present version of the index to our sample of Hungarian research organizations, in comparison with the previous versions of the index.

## Results and discussion

In order to explore the performance of the modified version of the Stirling index on our sample, we ran the calculation of research performance diversity for Hungarian Research Organizations both with the original formula of Rafols, Leydesdorff and Meyer, which uses cosine distance as the distance parameter, and with the new formula suggested here, which incorporates the weighted path length measure (again, calculations were implemented in the R statistical software). This resulted in two series of diversity values, and provided two different rankings of HROs, according to the diversity of their research profile. Of particular interest are (1) the position of the organizations along the two rankings, and (2) the comparison of the two measures applied.

Fig 1. shows the position of each HRO according to the two measures in a comparative manner. On this diagram, the ranking provided by the original version of the index is plotted against the ranking yielded by the modified version (introduced by us above): the horizontal and the vertical axes show the ordering obtained by the cosine similarity version and the weighted path length version, respectively. Ranks were assigned to insitutes in a reverse order: the HRO with the maximal diversity has the maximal rank number (27), while the least diverse HRO is indicated with the minimal rank number (1). The dotted line represents the points where the two rankings would be identical. The diversity values underlying the rank-orderings are presented in Table 3 of the Appendix.

As seen from the scatterplot, the two versions of the Stirling index are, in general, pretty much in agreement (for important differences, see below). The Spearman rank correlation of the two indices is 0.92 (statistically significant). An overall summary is as follows. On the top of both lists are Hungarian research universities (CORV, NYME, ELTE, etc.) together with the set of research institutions of the Hungarian Academy of Sciences (i.e., MTA as a whole). The middle range of both lists is shared by universities with a concentrated educational profile (BME, SZIE) as well as some smaller, non-profit institutions (COLBUD, BAY). Further down on both rankings we find organizations pursuing research primarily in the life sciences, such as pharmaceutical firms (EGIS, RICHTER) or hospitals (HEIM), but the oldest and largest Hungarian medical university also belongs to this group (SOTE). At the bottom of both lists, mostly governmental research institutions with a relatively narrow research area are found (ATK, MAFI, ONK, PSYNEU).

Although the overall picture thus gives little surprise, a closer look on the results reveals considerable features of both the institutional research profiles and the content of the proposed diversity measure. As to the former, consider the position of universities on this plot. One might naively argue that universities with a pre-assigned, relatively broad educational profile naturally produce a more diverse research portfolio than do universities with a more narrowly defined mission. However, having a number of faculties and providing education in different areas does not, by itself, entail *publishing* or *being cited* in all of those areas. Most striking is the comparison between the Corvinus University (CORV) and the Szent István University (SZIE). The educational profile of both institutions covers primarily the fields of economics and social studies, applied life sciences, agri-, horticulture and engineering, and these occur in different proportions and historical settings. Still, CORV is the absolute winner of both rankings (at the top right corner on the plot), showing the same (maximal) rank by both indices, while SZIE is in the middle range, and again, with almost identical twin ranks. Also notable is the relative position of MTA and ELTE, the latter widely held to be the most prominent Hungarian research university: according to both lists, the diversity of the research output of these two organizations goes side by side, suggesting that the performance of a Hungarian top university and the governmental network of research institutions, jointly covering the entire science system, are quite comparable with respect to publication diversity.

PTE, a university with a broad educational profile integrating institutionally more areas than ELTE, is placed at a much lower rank by both measures, having the same position in the middle range of the two orderings (its datapoint lies on the dotted line of the plot). At the same time, NYME, a university generally considered to be targeted at agriculture and ecology, is favoured by each index (though less by the proposed new one) above ELTE or MTA as our bottomline examples.

Facing the performance of the measure(s) under study, it is intriguing to explore what features of competence maps the calculated values and ranks reflect, in order to better understand now the aspect of „diversity" captured by the proposed index. Figure 2, in addition to the scatterplot, shows the change in the ranking of each HRO resulting by the substitution of the original version of the Stirling index by the proposed new one. Since, conceptually, the proposed new index differs from the original one by importing the notion of path length, a ranking yielded by a third index, which emphasizes the effect of using raw path lengths is also shown in the third column. Note that the index used in the third column is virtually identical to the version introduced by Rafols and Meyer 2010), containing the (unweighted) shortest path length for the distance measure (expressed as the number of edges). Indeed, since there is less difference between the two path length based rankings, than between either of them and the original index (the first column on the plot), the re-configuration of the list can be expected to be attributable to the use of a path length-like distance parameter.

Instantly observable by this visualization is that, in spite of the high general agreement between the cosine distance version and the weighted path length version, local (and less local) changes in the rank of institutions occur when switching the index. Most striking is the increase in the rank of CEU (Central European University), positioned in the middle range by the original Stirling index, but nominated as the „second best" by the proposed new index. A similar slope of the increment in rank is exhibited by OEP, a governmental health institution, and the Catholic university PAZM, while a more modest, still salient upgrade can be attributed to the college NYÍR. Looking at structural features of the research performance underlying these changes, one can find some common characteristics of the competence maps for these organizations. Their maps are, notably, not densely populated by active Subject Categories: instead, one can identify a handful of active areas, individual SCs or groups of closely related SCs, that occupy otherwise distant areas of the network. In the case of CEU, it is the extreme regions, or „poles" of the network where most of the recognizable publication activity is present (such as in SCs belonging to „Economics, Politics and Geography" and „Engineering", respectively), while the areas falling in between those distant regions are represented by a much more modest production. Even more telling are the cases of PAZM, OEP and NYIR (a Catholic university, a governmental body, and a College, respectively), where on the competence maps a couple of distant regions exhibit themselves as quite active, while the rest of the map is otherwise empty, that is, no activity can be observed in all other regions. Conversely, for HROs where the map is densely populated, or, to put it differently, where there are no wide „gaps" between the active areas, one experiences a certain degree of downgrading by the newly proposed index relative to the original. Though remaining at the top league of the list, ELTE and MTA are placed lower in the second ordering by the weighted path length version: evaluating this change against the maps of these HROs, it turns out that these institutions are very diverse in terms of the number and the distribution of subject categories, however, the heavily contributing SCs populate a broad but rather continuous region of the network, thus reducing the overall distance between the active areas of research.

Summarizing the observations described above, we might claim that the meaning of „diversity" measured by the proposed new version of the Stirling index differs, to some extent, from the usual measure. Favouring, as we have seen, activities in distant and

„unconnected" areas, the new index conveys the notion of how *polarized* a certain research portfolio is (where „unconnected" is but a metaphor here, expressing that the intermediate areas are inactive, the term does not refer to the structure of the network itself). As an effect of the index's emphasis on path length values, active contribution within a few distant areas/poles of the science system is scored higher by this concept, than evenly distributed effort in a high number of areas. As a consequence, we might call this measure a *polarization index*, rather than a diversity index, in the commonly accepted sense of the word.

We prefer to conceptualize this alternate semantics as reflecting on a different *type* or *mode* of research performance diversity, than the one emphasized in previous studies. This formulation paves the way to a promising work on a possible typology of research portfolios based on the *type of diversity* they exhibit. The result, namely the fact that the measure discussed above seems to recognize the degree of polarization, while the original version is more sensitive to evenly distributed research in different fields, suggests that the systematic exploration of such indices would reveal a *typology of diversity* itself, with respect to research output. Evaluating research from this perspective might provide a much deeper insight into the structure of research potentials of an institution, that can, in turn, support a more informed and more suitable assessment of the insitutional players of science.

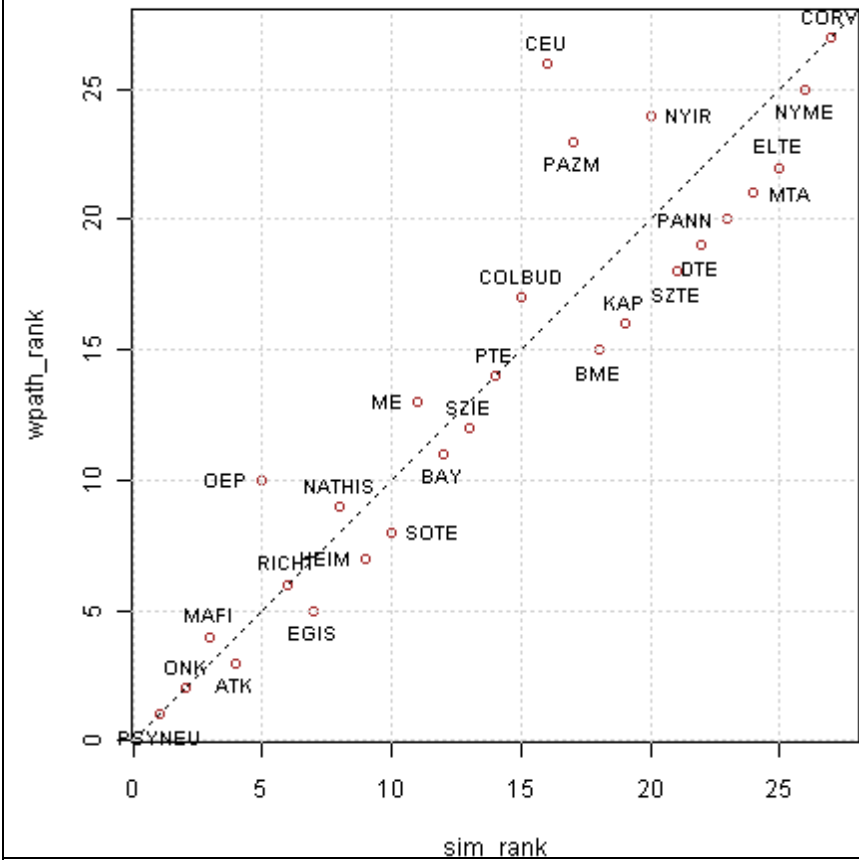

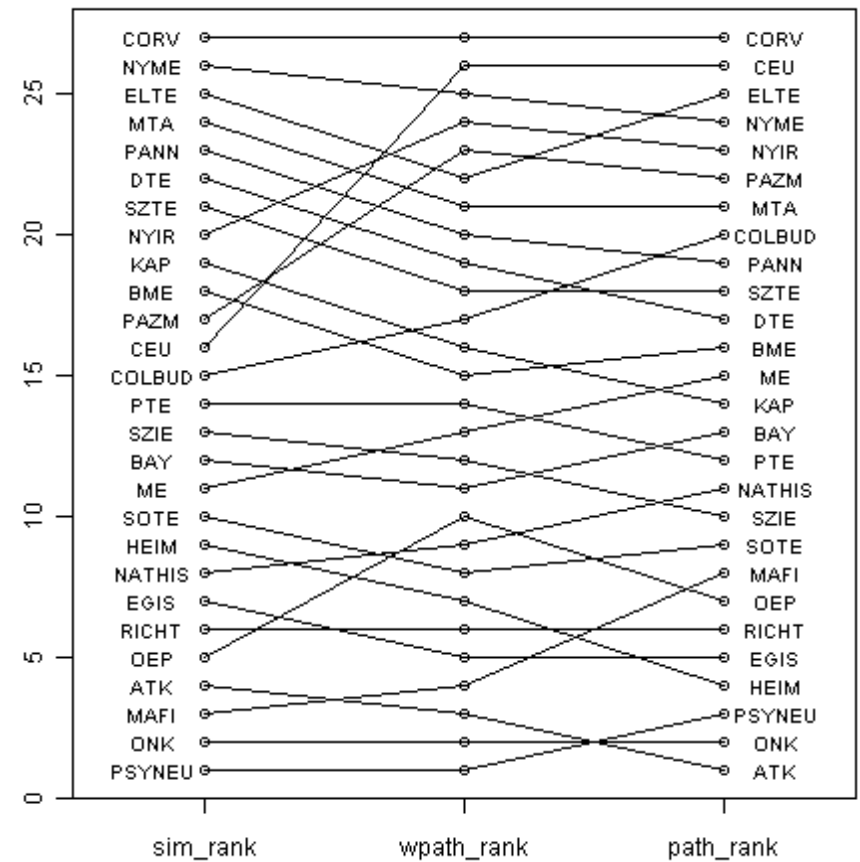

**Fig 1.** The diversity rank of Hungarian Research Organizations (HROs) using the weighted path length (wpath_rank) vs. the cosine distance measure

**Fig 2.** Changes in the diversity rank of HROs when applying three different distance measures: the cosine distance (sim_rank), the weighted path length (wpath_rank) and the (unweighted) path length (path_rank).

**Fig 3**. *Competence map of four HROs: the groups A)–B) and C)–D) reflect different types of research diversity. For color figures of the 27 major HROs, see http://hungarianscience.org*

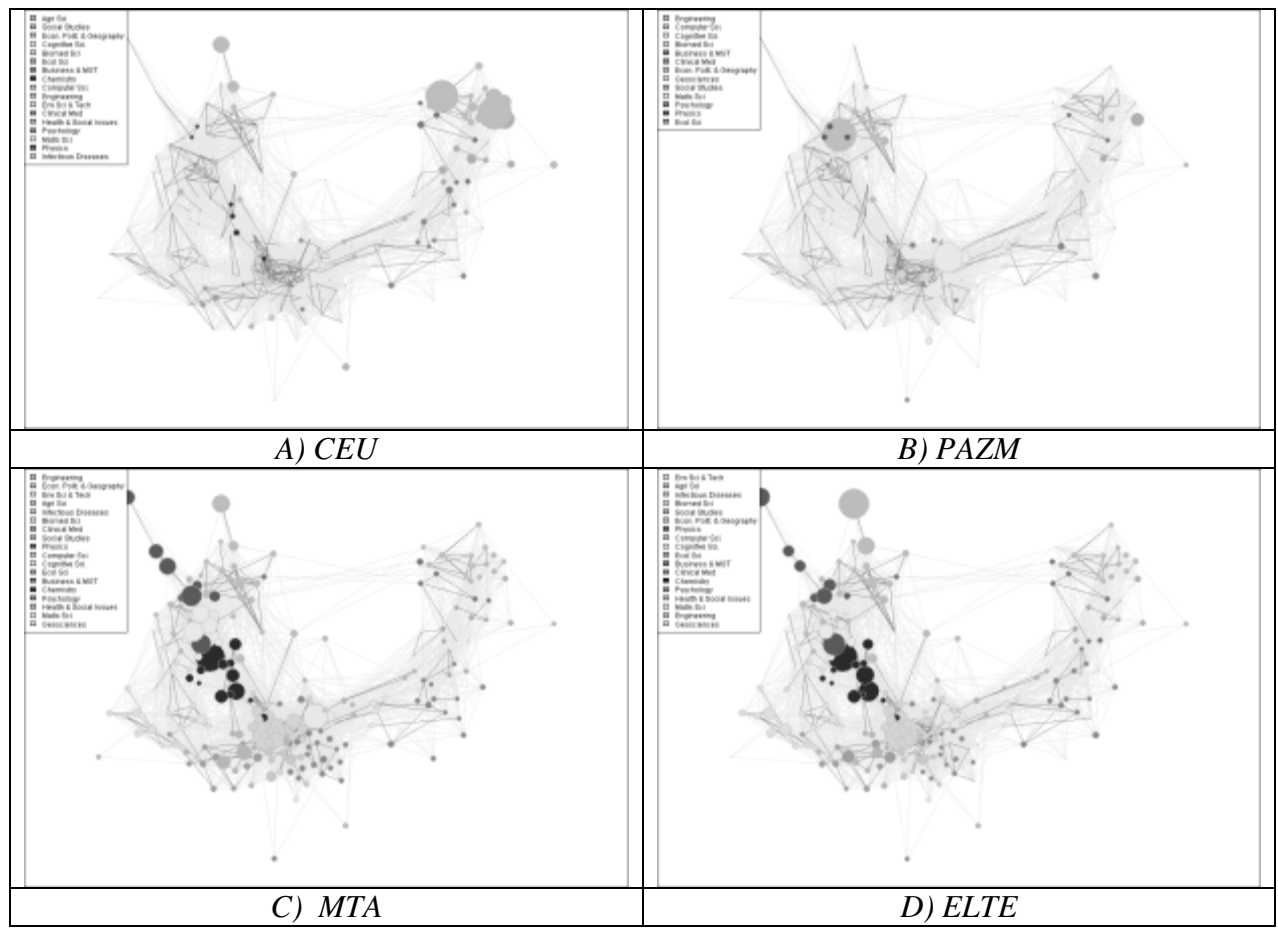

A) CEU     B) PAZM

C) MTA     D) ELTE

## Appendix

**Table 2**. *The list of major Hungarian Research Institutions (HROs)*

| Abbrev. | Institution |
|---|---|
| ATK | Research Institute for Animal Breeding and Nutrition |
| BAY | Bay Zoltan Foundation for Applied Research |
| BME | Budapest University of Technology & Economics |
| CEU | Central European University |
| COLBUD | Collegium Budapest Institutie for Advanced Study |
| CORV | Corvinus University Budapest |
| DTE | University of Debrecen |
| EGIS | EGIS Pharmaceutical Ltd |
| ELTE | Eötvös Loránd University |
| HEIM | Heim Pál Children's Hospital |
| KAP | University of Kaposvár |
| MAFI | Geological Institute of Hungary |
| ME | University of Miskolc |
| MTA | Hungarian Academy of Sciences |
| NATHIS | Hungarian Natural History Museum |
| NYIR | College of Nyiregyháza |
| NYME | University of West Hungary |
| OEP | National Health Insurance Fund Administration of Hungary |
| ONK | National Institute of Oncology |
| PANN | Pannon University |
| PAZM | Péter Pázmány Catholic University |
| PSYNEU | National Institute of Psychiatry & Neurology |
| PTE | University of Pécs |
| RICHT | Gedeon Richter Chemical Works Ltd |
| SOTE | Semmelweis University (of Medicine) |
| SZIE | Szent István University |
| SZTE | University of Szeged |

**Table 3.** Values obtained from applying the two versions of the Stirling index on the sample of HROs: div_sim and div_wpath stands for the cosine distance version and the weighted path length version, respectively. The list is ordered by the div_path values.

| | div_sim | div_wpath | | div_sim | div_wpath |
|---|---|---|---|---|---|
| CORV | 0,446 | 0,653 | KAP | 0,406 | 0,474 |
| CEU | 0,396 | 0,598 | BAY | 0,376 | 0,415 |
| ELTE | 0,431 | 0,553 | PTE | 0,389 | 0,445 |
| NYME | 0,433 | 0,563 | NATHIS | 0,343 | 0,413 |
| NYIR | 0,413 | 0,561 | SZIE | 0,380 | 0,444 |
| PAZM | 0,403 | 0,554 | SOTE | 0,361 | 0,383 |
| MTA | 0,428 | 0,529 | MAFI | 0,289 | 0,347 |
| COLBUD | 0,395 | 0,490 | OEP | 0,311 | 0,414 |
| PANN | 0,422 | 0,514 | RICHT | 0,339 | 0,356 |
| SZTE | 0,415 | 0,500 | EGIS | 0,340 | 0,355 |
| DTE | 0,415 | 0,502 | HEIM | 0,358 | 0,372 |
| BME | 0,404 | 0,458 | PSYNEU | 0,242 | 0,243 |
| ME | 0,367 | 0,445 | ONK | 0,253 | 0,256 |
| | | | ATK | 0,292 | 0,299 |

**Fig 4**. *Distribution of shortest path lengths for the map of science used as basemap (left: unweighted path lengths measured by the number of edges; right: weighted path lengths measured by the sum of edge weights).*

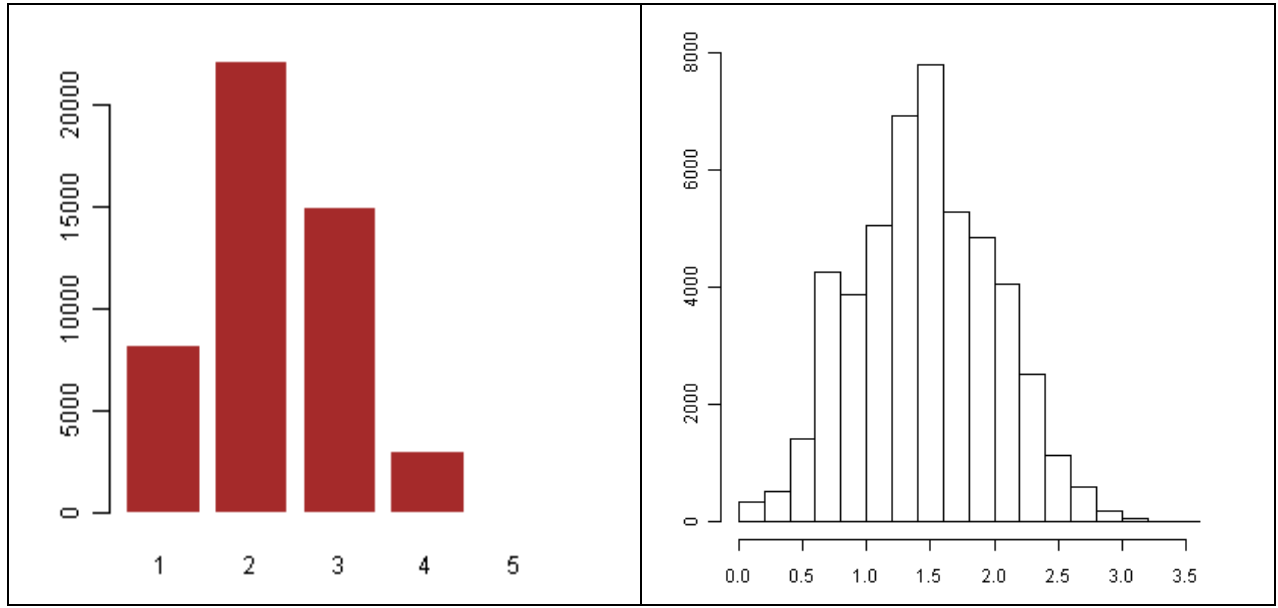

*Conceptual relations of the distance measures of the proposed and the popular versions of the Stirling index*

The relation between the proposed distance measure, the weighted path length $g_{ij}^W$, and the popular distance measure, the cosine distance is illustrated by Fig 5. On the graph, a configuration of SCs characteristic of the basemap is depicted: SCJ can be reached from SCI in two steps, and vice versa, as these are connected by SCK. The dotted line stands for an edge that is either nonexistent or omitted, bearing a weight below the chosen minimum of similarity values. On the lines edge weights are indicated.

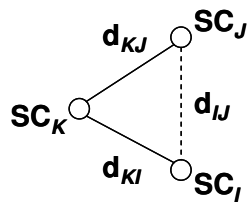

**Fig 5.** The configuration of three SCs in the science map

In this case, the two calculations would provide the following results.

(1) *Cosine distance*. Applying the cosine distance as the distance parameter means using directly the edge weights in the Stirling index. In the case shown below, the distance between $\mathbf{SC}_I$ and $\mathbf{SC}_J$ is

$$cos\left(\mathbf{SC}_I, \mathbf{SC}_J\right) = \mathbf{d}_{IJ} \, .$$

(2) *Weighted path length*. By our definition, the distance between $\mathbf{SC}_I$ and $\mathbf{SC}_J$ is the weighted (shortest) path length, which, assuming that the direct link between the two is omitted, yields

$$g^W(\mathbf{SC}_I, \mathbf{SC}_J) = \mathbf{d}_{KI} + \mathbf{d}_{KJ}.$$

Since the direct link connecting the two SCs is missing, the cosine *similarity* between them is zero (or set to zero by imposing the threshold), by which the cosine distance (1–cosine similarity) equals to 1. The sum of the other two weights could be either above, or below this value, since the cosine distance does not obey to the rule of triangle inequality (Korenius-Laurikkala-Juhola 2007), by which fact

$$\mathbf{d}_{KI} + \mathbf{d}_{KJ} \geq \mathbf{d}_{IJ} \text{ does not hold, therefore } cos(\mathbf{SC}_I, \mathbf{SC}_J) \leq \geq g^W(\mathbf{SC}_I, \mathbf{SC}_J)$$

For the diversity measure, these considerations have the consequence that using the weighted path length measure can make the diversity index much more sensible to the relatedness of Subject Categories in that it approximates otherwise uniformly treated distances (missing links) by different values. Theoretically, this also holds for the case where no threshold is used, and the calculation depends on the full weight matrix of the network, which means placing back some links to the system. Set aside the remaining set of missing links (the representing SCs not related at all), the path length values might also differ from cosine similarity values for below-threshold connected SCs as well, mirroring a different kind of connection between them.